\documentclass[twocolumn]{aastex631}

\graphicspath{{./}{figures/}}
\begin{document}

\title{A multi-faceted view of the X-ray spectral variability in Seyfert galaxy Ark 120}

\author{Lu-Xin Ren}

\author{Jun-Xian Wang}

\author{Jia-Lai Kang}

\affiliation{\rm CAS Key Laboratory for Research in Galaxies and Cosmology, Department of Astronomy, University of Science and Technology of China, Hefei, Anhui 230026, China; \url{renluxin@mail.ustc.edu.cn}, \url{jxw@ustc.edu.cn}, \url{ericofk@mail.ustc.edu.cn}}
\affiliation{\rm School of Astronomy and Space Science, University of Science and Technology of China, Hefei 230026, China}

\begin{abstract}
Utilizing a range of techniques including multi-band light curves, softness ratio analysis, structure functions, rms spectra, cross-correlation functions, and ratios of spectra from different intervals, we present a comprehensive study of the complex X-ray spectral variability in Seyfert 1 galaxy Ark 120, through re-analyzing its six XMM-Newton observations taken between 2003 and 2014. We find a clear ``softer-when-brighter" trend in the 2--10 keV power-law component over long timescales, with this trend being timescale dependent, as it is much weaker on shorter timescales, similar to that previously detected in NGC 4051.  
Notably, a rare ``harder-when-brighter" trend is observed during one exposure, indicating dynamic changes in the spectral variability behavior of the power-law component. This exceptional exposure, with the spectral variability indeed marked by a power-law pivoting at an unusually low energy of $\sim$ 2 keV, suggests intricate variations in the thermal Comptonization processes within the corona. Furthermore, when the data below 2 keV are included, we identify that the soft excess component adds significant complexity to the spectral variability, such as evidenced by a transition from ``harder-when-brighter” to ``softer-when-brighter” during another single exposure.
Such extra complexity arises because the variability of the soft excess sometimes follows and sometimes does not follow the changes in the power-law component.
Our findings underscore the necessity of applying multiple analytic techniques to fully capture the multifaceted spectral variability of AGNs.

\end{abstract}


\section{Introduction} \label{sec:intro}

Active Galactic Nuclei (AGNs), the dominant X-ray emitters in the extragalactic sky, are among the most luminous and variable sources in the universe, powered by the accretion of matter onto supermassive black holes (SMBHs) at their centers. It is widely accepted that the primary X-ray emission in AGNs (i.e., the power-law spectrum with a high-energy cutoff, \citealt{Zdziarski_1995,Ricci_2011,Tortosa_2018, Kang2020, Kang2022}) originates from inverse Compton scattering of optical and ultraviolet seed photons from the accretion disk by high-energy electrons in a compact, hot corona near the SMBH 
\citep[e.g.][]{Galeev_1979,Haardt_1991,Haardt_1993}. 
However, the physical nature of the corona remains poorly known.  
 
X-ray spectral variability in AGNs could provide essential insights into the nature of the X-ray corona, and the mechanisms driving the observed variability. For instance, a prominent feature of X-ray spectral variability is that in AGNs with intermediate to high Eddington ratios, a ``softer-when-brighter” behavior, i.e., the power-law spectrum steepens when the X-ray flux increases, is generally seen \citep[e.g.][]{Markowitz_2004,Sobolewska_2009,Soldi_2014}. Although it has long been suspected that the corona could be cooled down when the X-ray flux increases, resulting in a softer spectrum, the underlying mechanisms behind this behavior remain poorly understood. It is particularly puzzling considering the coronal temperature generally tends to increase with X-ray flux (\citealt{Keek2016,Zhang2018,Kang_2021,Pal2023}, but see \citealt{Barua2020,Wilkins2022,Masterson2022} for opposite examples), which contrarily would result in harder spectra at higher X-ray flux if the geometry/opacity of the corona remain unchanged. Other than changes of coronal temperature, non-static scenarios 
such as jet-like flaring corona \citep{Wilkins2015, Alston2020} or outflowing corona \citep{Liu2014} should  be involved. 
Meanwhile, there are also AGNs showing insignificant “softer-when-brighter” behavior, or even “harder-when-brighter” behavior (probably mostly at low Eddington ratios), suggesting different origins in their X-ray emission or variability \citep[e.g.][]{Emmanoulopoulos_2012,Connolly_2016}. 
Remarkably, \cite{Wu_2020} revealed that an empirical ``softer-when-brighter” trend is insufficient to explain the observed spectral variability in NGC 4051, a Seyfert galaxy with an Eddington ratio of $\sim$ 0.2 \citep{Yuan2021}, and the ``softer-when-brighter” behavior is timescale dependent. The spectral variability track for a single source may also vary between observations \citep[e.g.][]{Sarma2015}. These facts further demonstrate the complexity of the underlying physical processes.  

In addition to the power-law continuum, soft X-ray excess--significant surplus in the soft X-ray ($ \textless$2 keV) range--has been observed in many sources \citep[e.g.][]{Arnaud1985,Gierlinski2004,Piconcelli2005,Crummy2006,Bianchi2009}. The nature of the soft excess is also unclear,  with two leading theories being considered. One theory proposes that the soft excess originates from Comptonized emission by an extra ``warm corona"  \citep[e.g.][]{Done2012,Petrucci_2013}. The other theory suggests that the soft excess is due to relativistically blurred reflection from the inner accretion disk, irradiated by the hot corona \citep[e.g.][]{Ross1993,Crummy2006}. 
The soft excess component can affect the observed X-ray spectral variability, and thus shall also be taken into account while analyzing the spectral variability. In deed, spectral variability can be used to probe the origin of the soft excess \citep[e.g.][]{Mehdipour_2011, Ponti_2012, Jin2013,Nandi2021, Jin_2021}.

Technically, there are various approaches to reveal the X-ray spectral variability. Directly comparing the spectra from different epochs is the most straightforward one. However, when analyzing rapid spectral variability within individual exposures, limited by the photon statistics, the hardness ratio (or softness ratio) is generally adopted to quantify the spectral slope and plotted against the count rate (or flux) to demonstrate the spectral variability \citep{Markowitz_2004,Sobolewska_2009}. The variation of the spectral slope or hardness ratio (termed as color variation in UV/optical studies) could be measured as a function of the timescale \citep{Sun2014,Zhu2016,Wu_2020}. Meanwhile, the rms spectrum which measures the X-ray variability amplitude as a function of energy \citep{Vaughan_2003,Sesar_2007,Zuo_2012} is also widely adopted. The rms spectrum could also be obtained for variability at different ranges of frequencies/timescales \citep{Middleton_2009, Uttley_2014, Hu_2022}. 
Assuming the variability at different X-ray energies are perfectly coordinated, higher variability amplitude at lower energy is equivalent to ``softer-when-brighter" and vice versa. Consequently, the coordination between variability at various energies \citep{McHardy_2004, Epitropakis_2017} shall also be examined, such as with Cross-correlation Function (CCF).

However, though the X-ray spectral variability in AGNs has been widely studied in literature, generally only a single (or a limited few) of the above approaches was adopted in individual studies, hindering full understanding of the spectral variability behaviors. Furthermore, the timescale dependence of the ``softer-when-brighter" trend has only been in studied in one source \citep[NGC 4051,][]{Wu_2020}.

Ark 120 is a Seyfert 1 galaxy at redshift of 0.0323, with an estimated central supermassive black hole mass of $\sim$ 1.5 $\times$ 10$^8$ M$_\odot$ \citep{Peterson2004}, and a low Eddington ratio of $\sim$ 0.05 \citep{Vasudevan2007}. It is considered as a ``bare" nucleus, i.e., without any UV/X-ray absorption detected \citep{Crenshaw2001,Vaughan2004}, thus an ideal target to probe the intrinsic X-ray spectral variability. 
Ark 120 has been observed six times by XMM-Newton \citep{Jansen_2001}, each with a duration $>$ 100 ks. These XMM-Newtonexposures have been  presented and investigated in a number of studies \citep{Vaughan2004,Matt2014,Mallick2017,Porquet2018, Lobban2018,Nandi2021}.
In this work, we re-analyze the XMM-Newtonexposures using all the aforementioned approaches to uncover the intriguing yet complex X-ray spectral variability behaviors in Ark 120, most of which have not been previously discovered or adequately interpreted in the literature.
We further demonstrate that combining these various approaches is essential, as any single approach alone would be insufficient to fully describe the complicated spectral variability we present in this work. This paper is organized as follow. We present in \S\ref{sec:data} the XMM-Newtondata we adopted and the data reduction process. In \S\ref{sec:analysis} we reveal the spectral variability using various approaches, followed by discussion in \S\ref{sec:discussion}.

\section{The XMM-NewtonData and Reduction} \label{sec:data}

\begin{table}
	\centering
	\caption{Six XMM-Newton exposures of Ark 120.}\label{tab:data}
	\begin{tabular}{ccc} 
\hline
Obs ID & start time & duration/net-exp (ks) \\
\hline
0147190101 & 2003-08-24 05:35:43 & 112/54 \\
\hline
0693781501 & 2013-02-18 11:39:53 & 130/79 \\
\hline
0721600201 & 2014-03-18 08:35:23 & 133/78 \\
\hline
0721600301 & 2014-03-20 08:41:22 & 132/84 \\
\hline
0721600401 & 2014-03-22 08:07:51 & 133/77 \\
\hline
0721600501 & 2014-03-24 07:59:53 & 133/79 \\
\hline
\end{tabular}
\end{table}

\par
\begin{figure}
\centering
\includegraphics[width=0.48\textwidth]{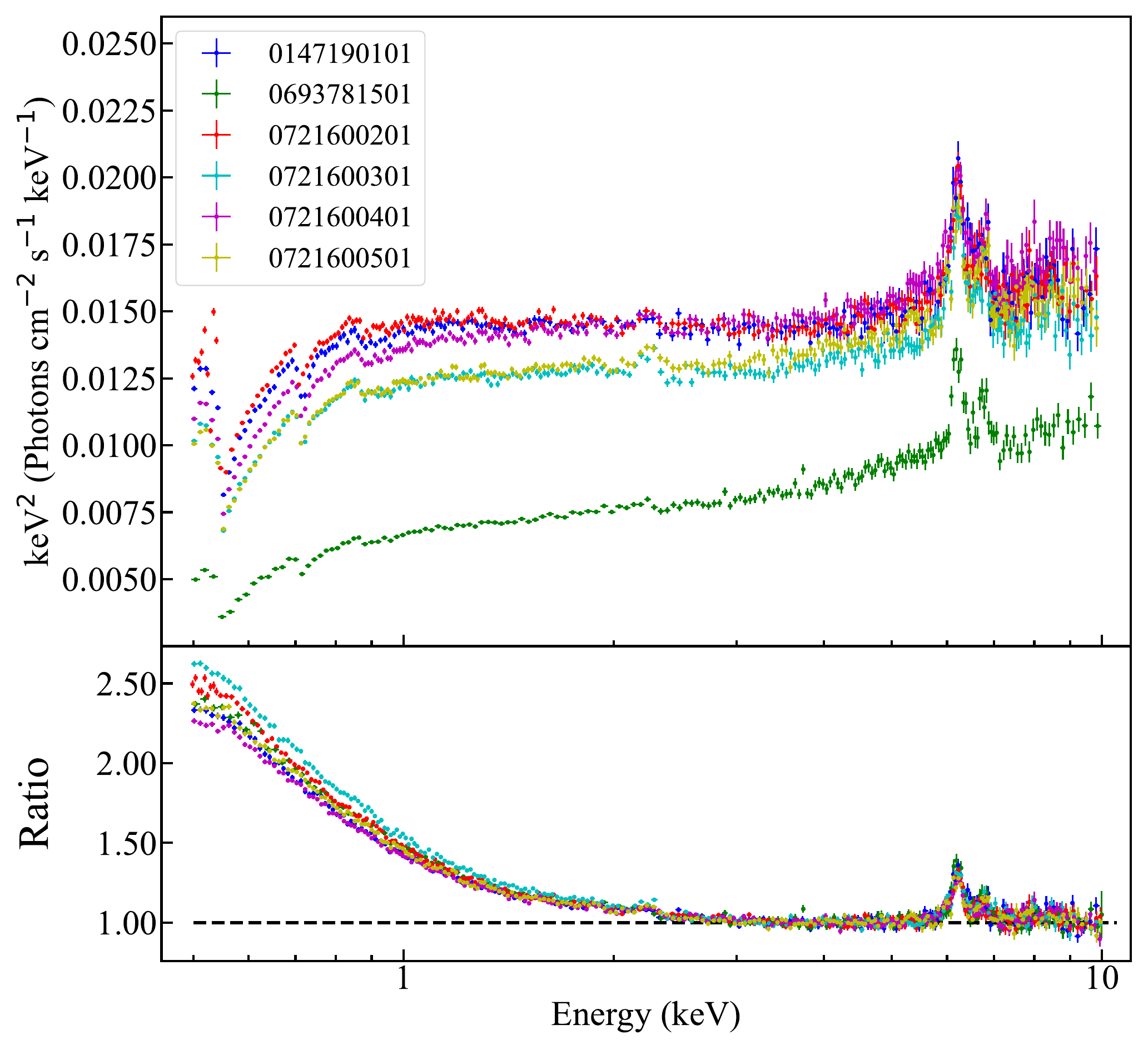} 
\caption{\label{fig:unfolded} Upper: the unfolded spectra (see below for the model adopted to derive the unfolded spectra) of six EPIC-pn observations of Ark 120. The spectra have been rebinned for display purpose.  Note though the unfolded spectra could be model dependent, such an effect is negligible while displaying the overall spectra profiles \cite[see also][]{Zhang2018}. Lower: the data to mode ratio plot, where the best-fit model is a single power-law absorbed by Galactic ISM \citep[$tbabs$,][]{Wilms_2000} with column density fixed at $N_{\rm H}$ = 1.4 $\times$ 10$^{21}$ cm$^{-2}$, yielded through fitting the spectra within 3.0--5.0 keV and 8.0--10.0 keV.}
\end{figure}

\begin{figure*}
\centering
\includegraphics[width=1.0\textwidth]{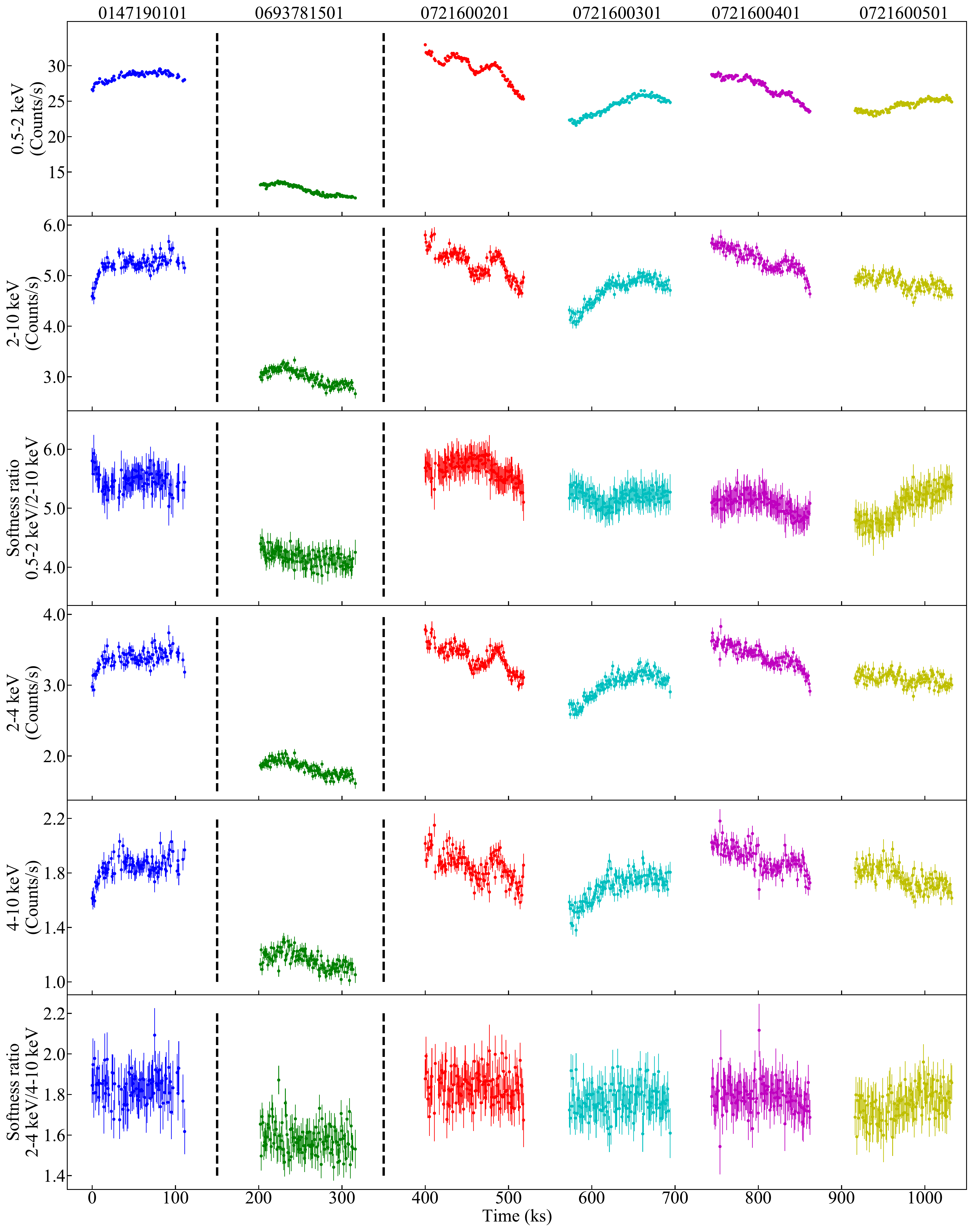} 
\caption{\label{fig:6lc} The X-ray light curves and softness ratios of the 6 EPIC-pn observations of Ark 120. The vertical dashed lines mark the discontinuities in x-axis because of too long gaps between exposures. }
\end{figure*}

 So far, Ark 120 has been observed six times by XMM-Newton.
Although these XMM-Newtonobservations have been well studied in literature \citep[e.g.][]{Vaughan2004,Matt2014,Mallick2017,Porquet2018,Nandi2021}, we provide, for the convenience of the readers, a summary of the six XMM-Newtonexposures in  Table \ref{tab:data}, their corresponding spectra in Fig. \ref{fig:unfolded} and light curves in Fig. \ref{fig:6lc}.

In this work we focus on data obtained with the EPIC-pn detector \citep{Struder_2001}, operated in small window mode for all six observations. Except for the second exposure (Obs ID 0693781501, which used the medium filter), all other exposures were obtained with the EPIC thin filter. 

We use the XMM-Newton Science Analysis System (SAS, version 20.0.0) and the current Calibration Files (CCF 3.13) to process the raw data. Following the procedures described in \cite{Kang2023}, we filter out high background intervals, and extract the source spectra within a circular region with a radius of 60\arcsec, and the background from nearby source-free regions. The pile-up effect is checked using SAS task $epatplot$ and is considered negligible for all exposures. The source spectra are further rebinned to have a minimum of 50 counts in each bin to enable $\chi^2$ statistics. 

To illustrate the soft X-ray excess clearly seen in literature, we fit the spectra within 3--10 keV (excluding the 5–8 keV range to avoid the influence of the broad Fe K$\alpha$ line) with $pexrav$ \citep{Magdziarz1995} absorbed by Galactic column density $N_{\rm H}$ = 1.4 $\times$ 10$^{21}$ cm$^{-2}$ \citep{Kalberla2005}.
Because the reflection fraction $R$ of $pexrav$ can not be well constrained through fitting XMM-Newtonspectra alone, we fix $R$ at $\sim$ 0.70 as measured through joint fitting NuSTAR and XMM-Newtonspectra after applying the inter-instrument calibration correction\footnote{https://xmmweb.esac.esa.int/docs/documents/CAL-TN-0230-1-3.pdf}  \citep{Kang2023}. Performing extensive spectral fitting however is out of the scope of this work. 

We could see that the X-ray spectra are dominated by the power-law component above 2 keV, and with clear contribution from the soft X-ray excess below 2 keV. In this work, we then utilize the softness ratio defined as the count rate ratio of 2.0--4.0 keV and 4.0--10.0 keV\footnote{The contribution of the Fe K$\alpha$ line (fitted with a narrow and a broad Gaussian) to 4.0 -- 10.0 keV count rate is only $\sim$ 3.3\%, making its potential impact on the softness ratio analysis negligible, even if its variability differs from that of the underlying continuum. } to probe the spectral variation of the power-law component. We also explore the spectral variation using the count rate ratio of 0.5--2.0 keV and 2.0--10.0 keV, which could be affected by both the power-law component and the soft X-ray excess. 

We extract the background-subtracted light curves for 0.5--2 keV, 2--4 keV, 4--10 keV, 0.5--10 keV, and 2--10 keV band respectively (each with a time bin of 1000 s), using the $epiclccorr$ task and applying both relative and absolute
corrections to correct various effects that affect the detection efficiency and enable the direct use of count rates and count rate ratios between various bands for spectral variability analysis. Note the difference between the transmission of EPIC thin and medium filter is negligible for this work. 
In Fig. \ref{fig:6lc} we plot the EPIC-pn light curves and softness ratios to illustrate the flux and spectral variability of Ark 120. Significant flux variations in different bands are clearly seen in all exposures in each band. Meanwhile, variations in spectral slope (softness ratio) are also visible in most exposures, particularly between 0.5--2.0 keV and 2.0--10.0 keV. We further exposure the variations of the softness ratio in \S\ref{sec:specvariation} and the rms spectra \S\ref{sec:rms}.
Additionally, we observe that the variability in various bands is generally well-coordinated, with the exception of Obs ID: 0721600501, where the 0.5--2.0 keV and 2.0--10.0 keV variations are anti-coordinated. The variability coordination between different bands will be discussed in \S\ref{sec:coordination}. Two individual exposures with exceptional spectral variability behaviors will be further examined in \S\ref{sec:abnormal}.

\section{The X-ray spectral variability} \label{sec:analysis}

\subsection{The ``softer-when-brighter" trend}\label{sec:specvariation}

We first explore the ``softer-when-brighter"  diagram of Ark 120 through plotting two softness ratios (SR1: 2.0--4.0 keV/4.0--10.0 keV; SR2: 0.5--2.0 keV/2.0--10.0 keV) against the corresponding total count rates (2.0--10.0 keV, and 0.5--10.0 keV, respectively). As aforementioned, while SR1 is dominated by the spectral variability of the power-law continuum, the soft X-ray excess may contribute significantly to SR2. 

\subsubsection{2.0--4.0 keV/4.0--10.0 keV}\label{sec:PL}

\begin{figure}
\centering
\includegraphics[width=0.48\textwidth]{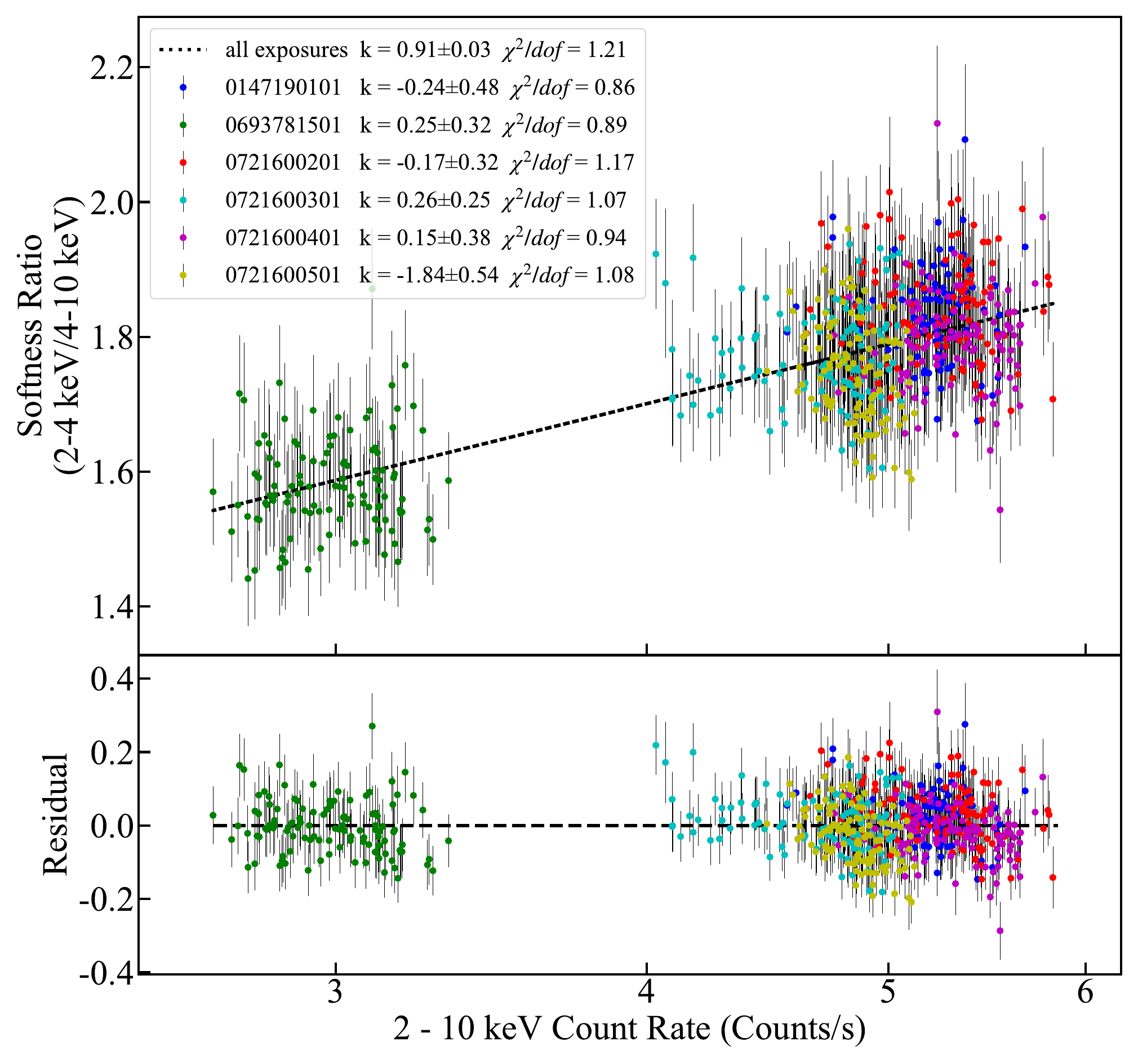}\\
\caption{\label{fig:SR1} The softness ratio (SR1: 2.0--4.0 keV/4.0--10.0 keV) versus 2.0--10.0 keV count rate for the 6 EPIC-pn exposures (color-coded) of Ark 120. The dashed line plots the best-fit linear regression line, with the yielded regression slope $k$ and the reduced $\chi^2$/dof labeled.  The regression slopes $k$ obtained using individual exposures only are also labelled. }
\end{figure}

\begin{figure}
\centering
\includegraphics[width=0.48\textwidth]{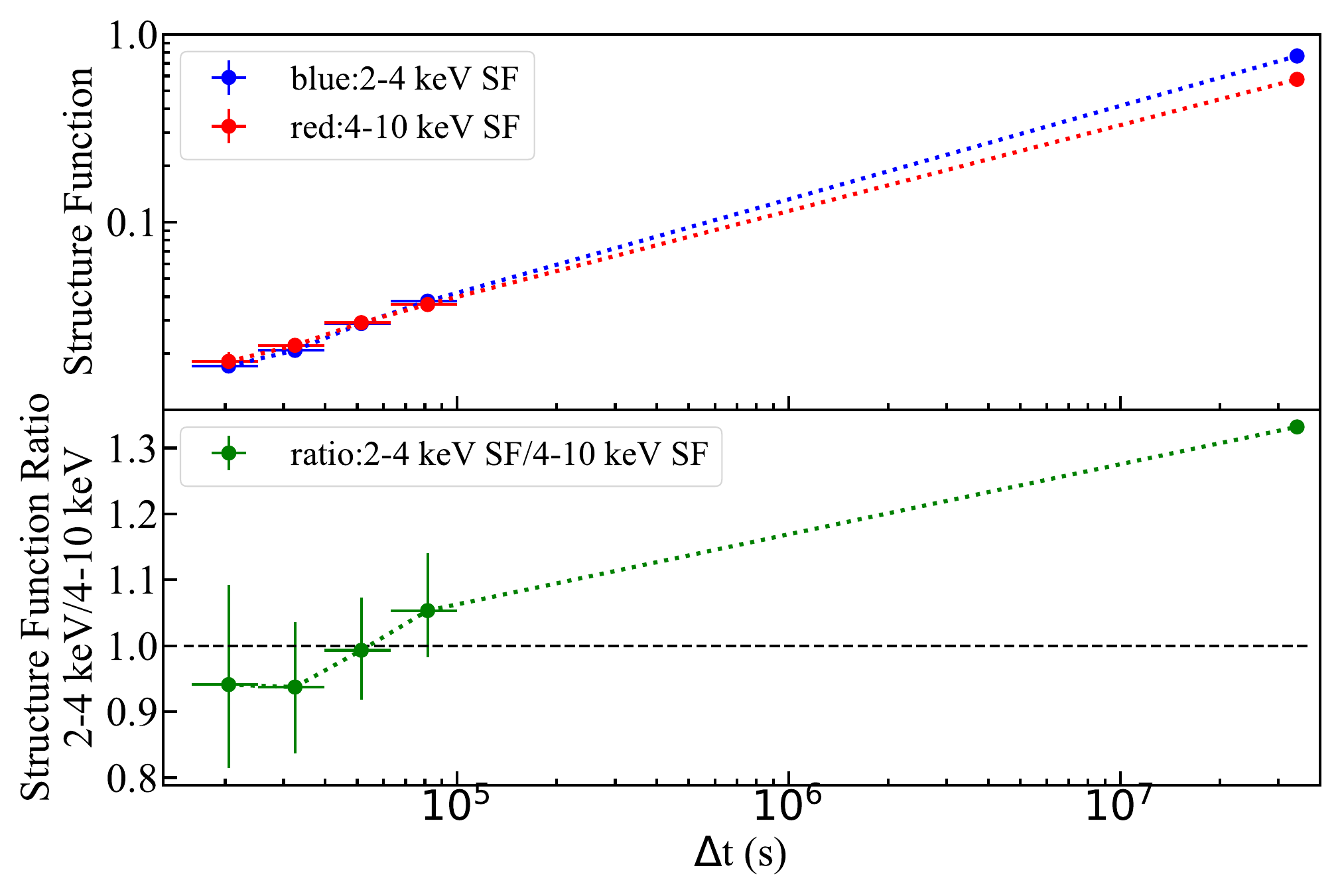}
\caption{\label{fig:SF} Upper: the structure function for the 2.0--4.0 keV and 4.0--10.0 keV band light curves of Ark 120. Lower: the ratio of two structure functions. }
\end{figure}  

In Fig. \ref{fig:SR1} (SR1 versus 2--10 keV count rate) we could see a clear ``softer-when-brighter" trend when putting all 6 observations together. The best-fit linear slope of the ``softer-when-brighter" trend is given in the plot. The yielded $\chi^2$/$dof$ = 1.21 suggests an empirical trend could be insufficient to describe the spectral variability, similar to that found for NGC 4051 \citep{Wu_2020}.

Indeed, if we zoom into individual exposures in Fig. \ref{fig:SR1}, we do not see a significant ``softer-when-brighter" trend in any of the individual exposures. Particularly, the last exposure (Obs ID: 0721600501) exhibits a contrary ``harder-when-brighter" trend (see the best-fit regression slopes $k$ marked in the plot). These facts also demonstrate the spectral variability within 2.0--10.0 keV of Ark 120 can not be fully described with a single empirical``softer-when-brighter" trend, and the absence of the trend in individual exposures suggests the spectral variability is timescale dependent as revealed in NGC 4051 \citep{Wu_2020}.

Following \cite{Wu_2020}, we further explore the timescale dependence of its spectral variability through plotting the ratio of the structure functions in two bands (see Fig. \ref{fig:SF}). The structure functions in 2.0--4.0 keV and 4.0--10.0 keV are obtained following the equation \citep[see also][]{diClemente1996,VandenBerk2004,Zhu2016}:

\begin{equation}
SF(\tau)=\sqrt{\frac{\pi}{2}<|Log(CR_{i})-Log(CR_{j})|>^{2}-<\sigma_{i}^{2}+\sigma_{j}^{2}>}
\end{equation}

in which $Log(CR_{i,j})$ represents the logarithmic count rates at any two epochs $i$, $j$ in the light curve, $\sigma_{i,j}$ the corresponding logarithmic statistical errors, and $\tau$ the lag between two epochs.
The uncertainties to the structure functions are obtained through bootstrapping the data points in the light curves \citep{Peterson2001}. According to \cite{Zhu2016}, the ratio of the two structure functions could quantify the ``softer-when-brighter" trend as a function of timescale.  From the lower panel of Fig. \ref{fig:SF} we could clearly see a prominent ``softer-when-brighter" trend (with SF ratio $>$ 1.0) at long term (timescale $>$ 10$^7$ s), and such a trend disappears at short term (timescale $<$ 10$^5$ s). We even see a marginal ``harder-when-brighter" trend (with SF ratio $<$ 1.0) at timescale $<$ 40 ks. Overall, the timescale dependence of the X-ray spectral variability we revealed in Ark 120 is similar with that seen in NGC 4051 \citep{Wu_2020}.

\subsubsection{0.5--2.0 keV/2.0--10.0 keV}\label{sec:SE}

\begin{figure}
\centering
\includegraphics[width=0.48\textwidth]{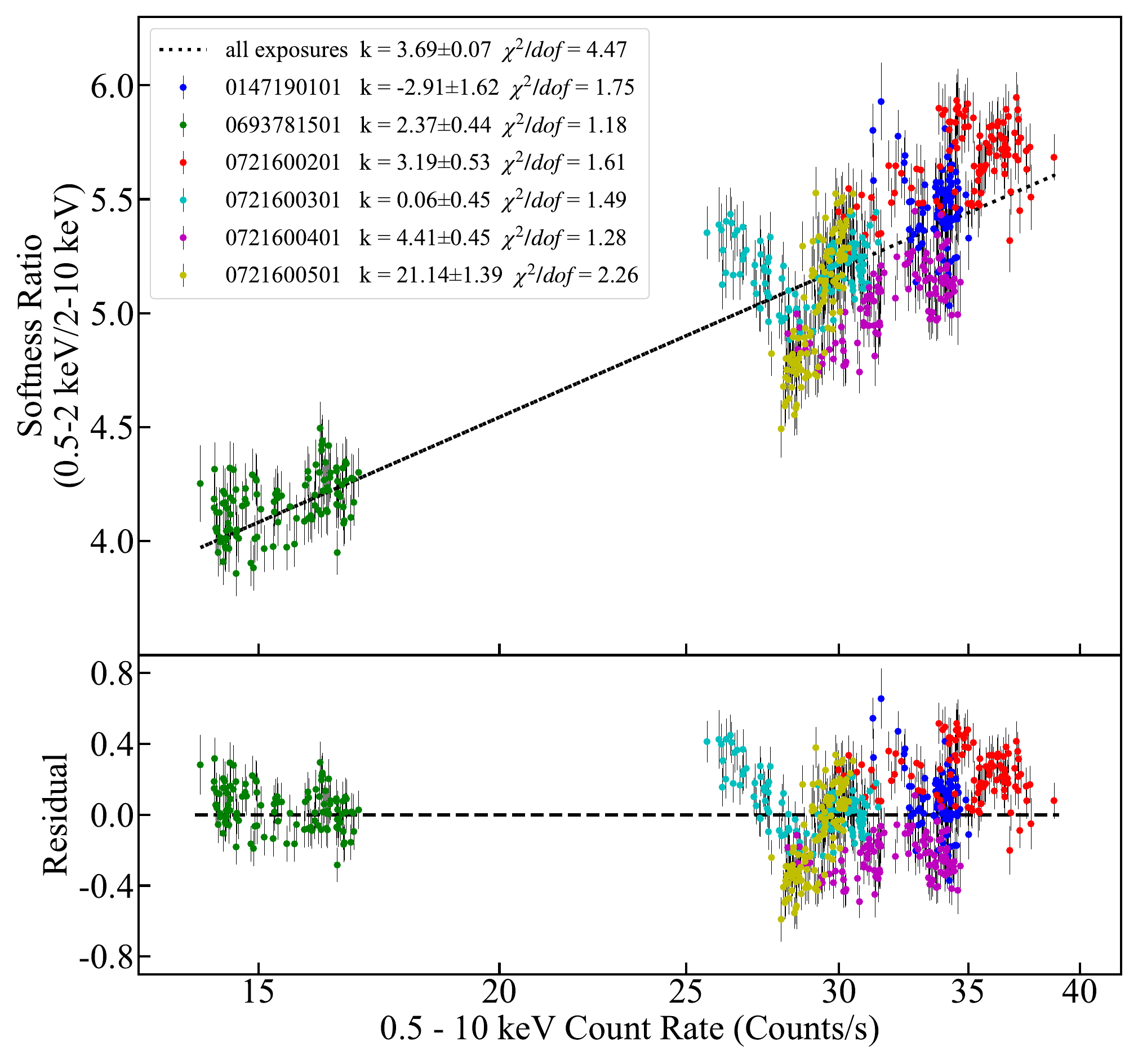}\\
\caption{\label{fig:SR2} Similar to Fig. \ref{fig:SR1} but for the softness ratio of 0.5--2.0 keV/2.0--10.0 keV (SR2) versus 0.5--10.0 keV count rate.}
\end{figure}  

We repeat the analysis in \S\ref{sec:PL} but now utilizing 0.5--2.0 keV and 2.0--10.0 keV light curves. Similarly, we see a clear ``softer-when-brighter" behavior ($k$ = 3.69$\pm$0.07) in the long term variation when considering all exposures together (see Fig. \ref{fig:SR2}, SR2 versus 0.5--10 keV count rate). 
We note \cite{Lobban2018} presented a figure quite similar to our Fig. \ref{fig:SR2} (see their Fig. 3), through plotting the hardness ratio (H+S/H-S, where H refers to 1--10 keV band count rate, and S refers to 0.3--1 keV band count rate) versus 0.3--10.0 keV count rate for the same 6 XMM-Newtonexposures of Ark 120, however without further interpretation except for pointing out the long term ``softer-when-brighter" behavior.

The similar long-term ``softer-when-brighter" behaviors seen in Fig. \ref{fig:SR1} and \ref{fig:SR2} suggest that while the hard X-ray power-law component follows a ``softer-when-brighter"  trend on timescale of years, the soft X-ray excess varies generally in coordination with the power-law on such a long timescale. The coordination between long-term variability of the power-law and the soft excess could also be seen in Fig. \ref{fig:unfolded} that the strength of the soft excess (see the data-to-model ratio plot in the lower panel) exhibits only weak variability between observations while the total count rate varies by a factor $>$ 2 (see Fig. \ref{fig:SR2}). 

Partially thanks to the much higher count rate in the 0.5--2.0 keV band, we see much more complicated spectral variations when we zoom into individual exposures. In the 1st exposure, we see a marginal ``harder-when-brighter" behavior ($k$ = -2.91 $\pm$ 1.62), and in the 2nd, 3rd, 5th exposure, we see ``softer-when-brighter" slopes similar to that derived from all exposures, but note the 3rd exposure appears significantly softer compared with the overall trend (the dashed line in Fig. \ref{fig:SR2}), and the 5th exposure appears systematically harder. The yielded $\chi^2/dof$ is 4.47 when putting all exposures together, and $>$ 1.2 for most individual exposures.  All these facts indicate a simple empirical ``softer-when-brighter" trend is far from sufficiently explaining the X-ray spectral variability within 0.5--10.0 keV.

More strikingly, we see a ``V" shape variation in the 4th exposure in Fig. \ref{fig:SR2}. Looking back to the light curves in Fig. \ref{fig:6lc}, we find in this exposure the softness ratio SR2 (0.5--2.0 keV/2.0--10.0 keV) switches from ``harder-when-brighter" during 0--50 ks of the exposure to ``softer-when-brighter"  after 50 ks. Furthermore, a much steeper slope ($k$ = 21.14) is seen for the 6th exposure.  The abnormal spectral variability in these two individual exposures will be further investigated in \S\ref{sec:abnormal}. 

\begin{figure}
\centering
\includegraphics[width=0.48\textwidth]{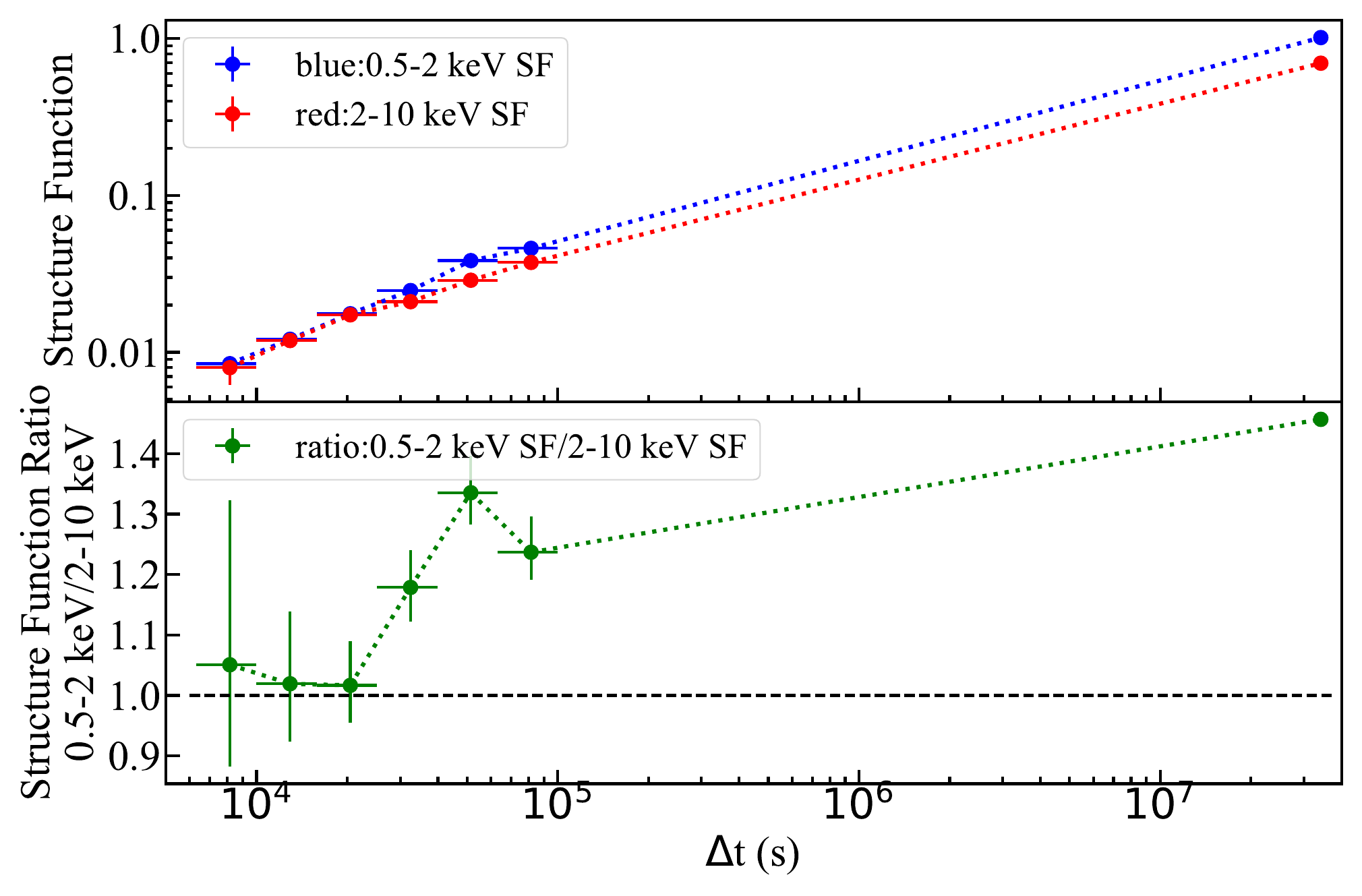}
\caption{\label{fig:SF2} Similar to Fig. \ref{fig:SF} but for 0.5--2.0 keV and 2.0--10.0 keV band.}
\end{figure}

Meanwhile, the timescale dependence of the ``softer-when-brighter" behavior (revealed utilizing 0.5--2 keV and 2--10 keV light curves) is shown in Fig. \ref{fig:SF2}. The timescale dependence is similar to that have shown in Fig. \ref{fig:SF}, indicating much weaker ``softer-when-brighter" behavior at shorter timescales.

\subsection{The rms spectra}\label{sec:rms}

\begin{figure}
\centering
\includegraphics[width=0.48\textwidth]{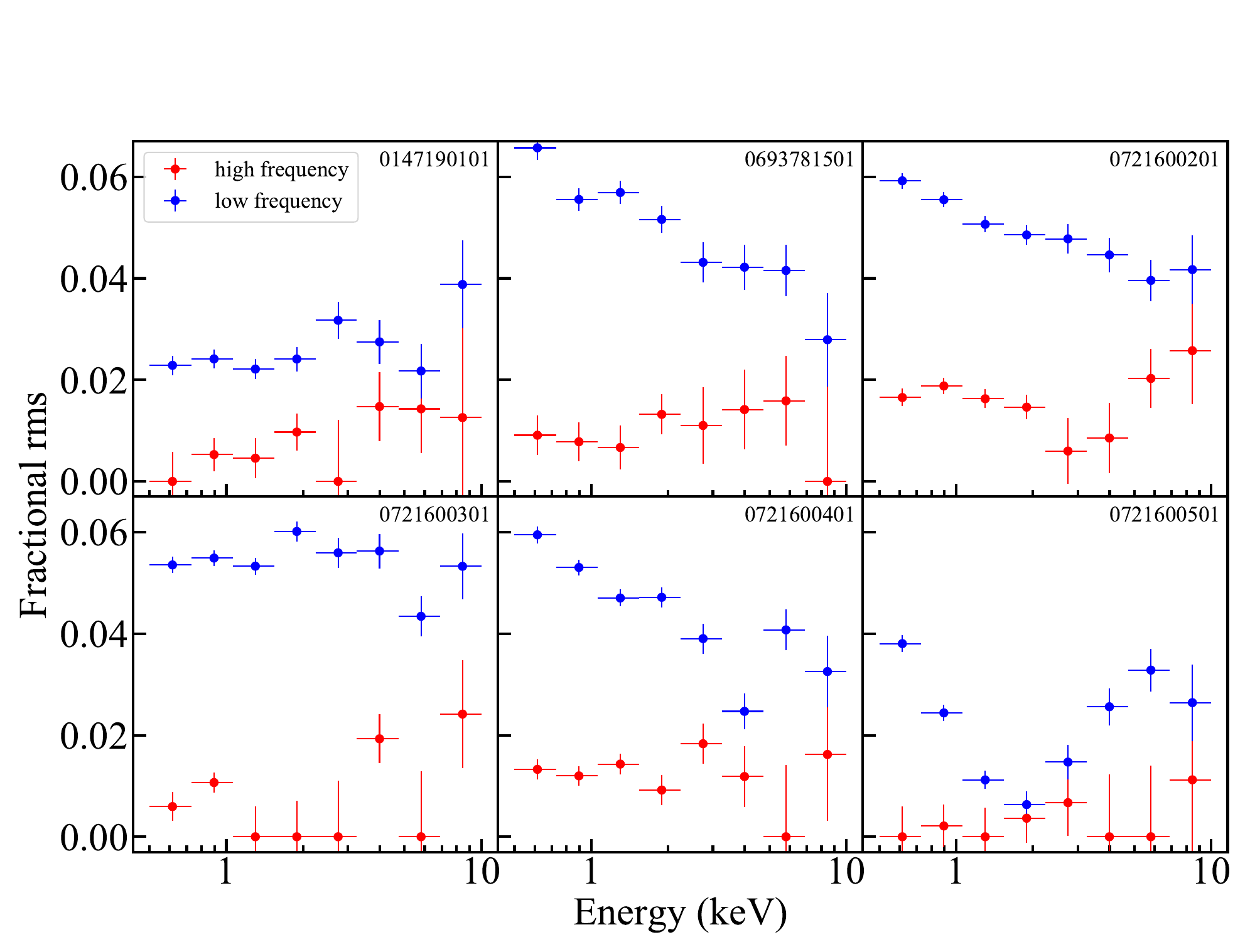}
\caption{\label{fig:rmsspectra} The high ($>$ 10$^{-4}$ Hz) and low frequency ($<$ 10$^{-4}$ Hz) rms spectra for the six XMM-Newtonexposures.  }
\end{figure}  

\par We also introduce root-mean-square (rms) spectra to examine the dependence of variability amplitude on energy. An observation with ``softer-when-brighter" trend will have a soft rms spectrum, i.e., stronger variability at lower energy, and vice versa. Similar to \S\ref{sec:data}, we extract the light curves within 8 energy bins between 0.5--10 keV. Following \citet{Arevalo_2008}, we calculate the frequency-resolved fractional rms and its error for each light curve, based on the power spectral density \citep[see also][for example]{Vaughan_2003, Poutanen_2008, Jin_2017, Hu_2022}. We calculate the rms spectra in two frequency ranges (high: $>$ 10$^{-4}$ Hz, and low: $<$  10$^{-4}$ Hz). 
The rms spectra within both frequency ranges are plotted for six observations, as shown in Fig. \ref{fig:rmsspectra}.

If we focus on the spectral range above 2 keV, because of the limited photon counts, we do not find clear energy dependence in the rms spectra, except for the 6th exposure where we see the low frequency rms spectra increases with increasing energy at $>$ 2 keV. This is consistent with the patterns we have shown in Fig. \ref{fig:SR1} that we do not see ``softer-when-brighter" trend in either of the exposures at $>$ 2 keV, but a ``harder-when-brighter" trend in the 6th exposure. We also see that at $>$ 2 keV, the high frequency rms spectra are marginally harder than the lower frequency ones in the 2nd, 3rd, 5th exposure. This also supports the diagram we have shown in Fig. \ref{fig:SF} that the ``softer-when-brighter" trend above 2 keV is timescale dependent. 

Once the spectral range below 2 keV is included, we see clearly soft low frequency rms spectra in the 2nd, 3rd, and 5th exposure, consistent with the ``softer-when-brighter" trend demonstrated in Fig. \ref{fig:SR2}. Again, the high frequency rms spectra appears harder than the low frequency ones in all but the 6th exposure, supporting the timescale dependency of the ``softer-when-brighter" trend shown in Fig. \ref{fig:SF2}.

The low frequency rms spectrum of the 6th exposure however appears exceptional, exhibiting a ``V" shape with minimum variability amplitude at $\sim$ 2.0 keV.  Such an abnormal rms spectrum was also noted by \cite{Mallick2017}, which presented the rms spectra ( 8--500 $\times$ 10$^{-6}$ Hz) for the 4 XMM-Newtonexposures obtained in 2014 (the 3rd -- 6th exposure in this work).  \cite{Mallick2017} modelled the rms spectra assuming the primary continuum ({\rm NTHCOMP}) is varying in photon index $\Gamma$ and normalization, and the soft excess is varying in luminosity. However, they did not address why the 6th exposure is exceptional by showing a ``V" shape rms spectrum. Moreover, the 6th exposure also exhibits abnormally steep ``softer-when-brighter" slope in Fig. \ref{fig:SR2}, for which we do not yet have an interpretation based on its rms spectra. 

Furthermore, during the 4th exposure we have seen transition from ``harder-when-brighter" to ``softer-when-brighter" (Fig.\ref{fig:SR2}). No clue on this transition could be extracted from the rms spectrum for the whole exposure. The two exposures (the 4th, and 6th) with abnormal spectral variability will be further discussed in \S\ref{sec:abnormal}.

\subsection{The cross correlation function}\label{sec:coordination}

As we attempted to understand the anomalous spectral variability of the 6th exposure (i.e., ``harder-when-brighter" in Fig. \ref{fig:SR1} but exceptionally steep ``softer-when-brighter" slope in Fig. \ref{fig:SR2}, and the abnormal ``V" shape rms spectrum in Fig. \ref{fig:rmsspectra}), we noticed its 0.5--2.0 keV and 2.0--10.0 keV light curves exhibit opposite variation trends. In other words, the variability in 0.5--2.0 keV is inversely coordinated with that in 2.0--10.0 keV. We further calculate the cross correlation function to assess the coordination between variability in different bands of the XMM-Newtonexposures. 

\begin{figure}
\centering
\includegraphics[width=0.5\textwidth]{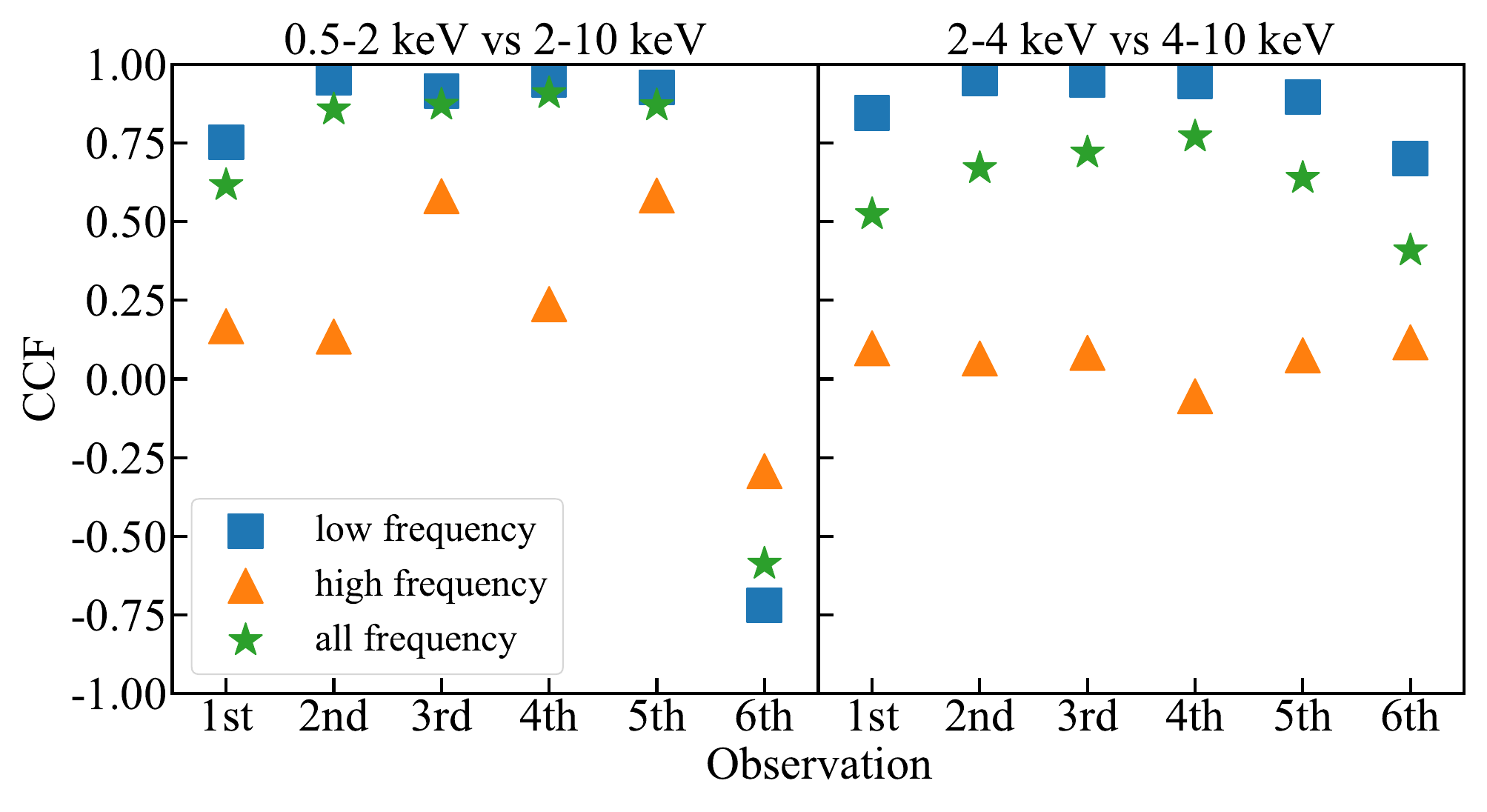}
\caption{\label{fig:CCF}  The cross correlation function values at zero lag between variations in various bands for all 6 XMM-Newtonexposures. The 1 ks bin light curves are adopted for the calculations, and the CCF values at different frequency ranges (high: $>$ 10$^{-4}$ Hz, and low: $<$  10$^{-4}$ Hz) are presented.}
\end{figure}

Using the 1 ks bin light curves, we calculate the CCF between variability in different bands. Since no clear lag is detected\footnote{\cite{Lobban2018} presented detailed frequency-dependent Fourier time lag analysis utilizing 100 s bin XMM-Newtonlight curves of Ark 120, and detected a high frequency soft X-ray lag of $\sim$ 900 s between 0.3--1.0 keV, and 1.0--4.0 keV. In this work we simply perform CCF analysis using 1 ks bin light curves to explore the coordination between variations in different bands. Studying the soft lag is beyond the scope of this work.}, we simply derive the CCF values at zero lag and plot them for all six XMM-Newtonexposures in Fig. \ref{fig:CCF}. The CCF values at zero log are also obtained for two frequency ranges (high: $>$ 10$^{-4}$ Hz, and low: $<$  10$^{-4}$ Hz).  

We find that, the low frequency variations are well coordinated (with CCF values close to 1.0) between 2--4 and 4--10 keV in all 6 exposures, and between 0.5--2 and 2--10 keV in all but the 6th exposure. The coordination between high frequency variations are generally weaker, likely due to Poisson noise of photon counts. The 6th exposure appears exceptional in the CCF analysis, that we see strong coordination between variations in 2--4 and 4--10 keV, but strong negative coordination between variations in 0.5--2 and 2--10 keV. The nature of such an exceptional variation is further explored below.

\subsection{The abnormal spectral variability in the 4th and 6th exposures}\label{sec:abnormal}

\begin{figure}
\centering
\includegraphics[width=0.5\textwidth]{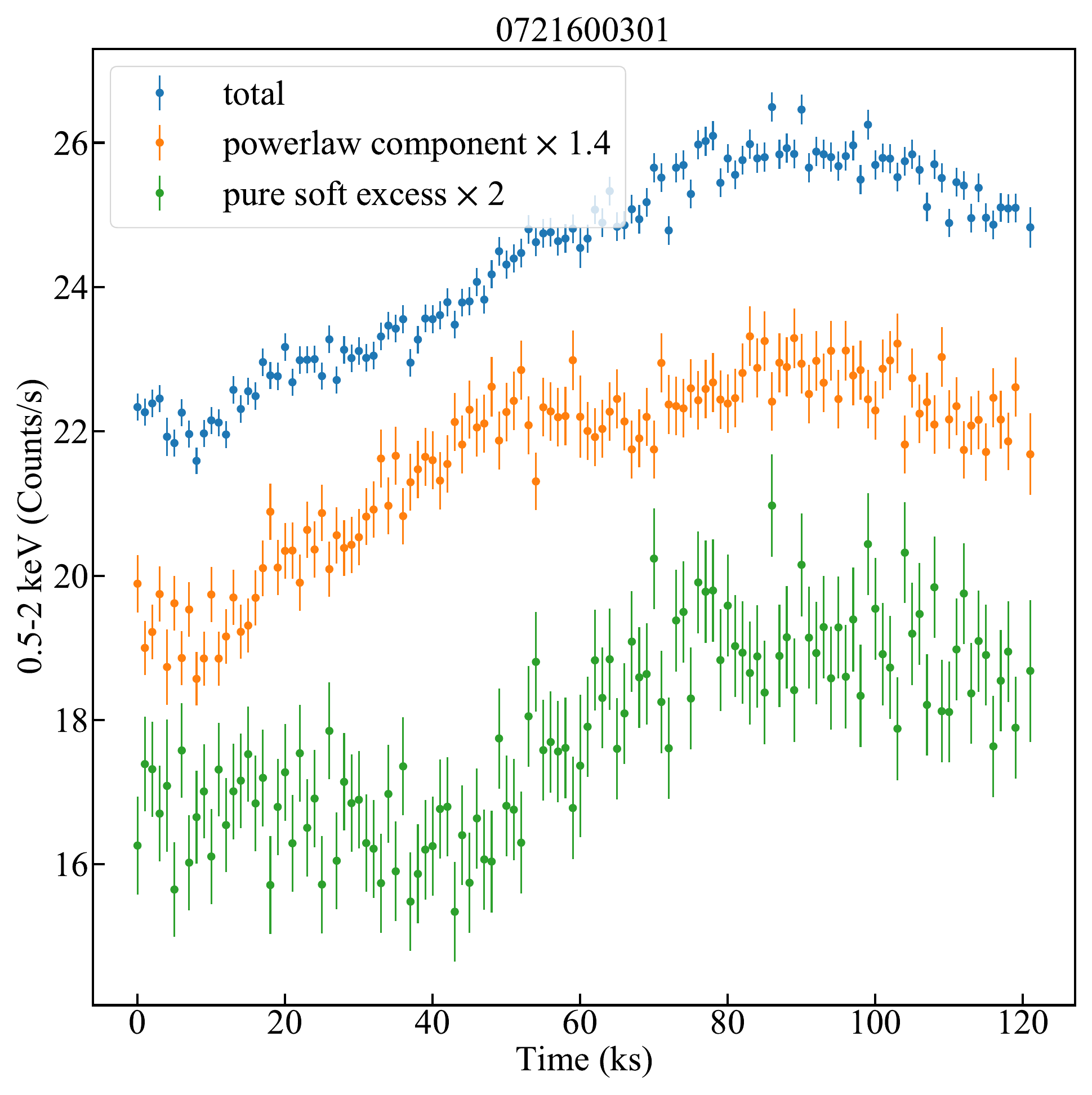}\\
\caption{\label{fig:4thLC} Blue: 0.5--2.0 keV light curve for the 4th exposure. Orange: the expected 0.5--2.0 keV light curve for the power-law component. Green: the light curve for the pure soft excess component (blue minus yellow). 
}
\end{figure}  

\begin{figure}
\centering
\includegraphics[width=0.5\textwidth]{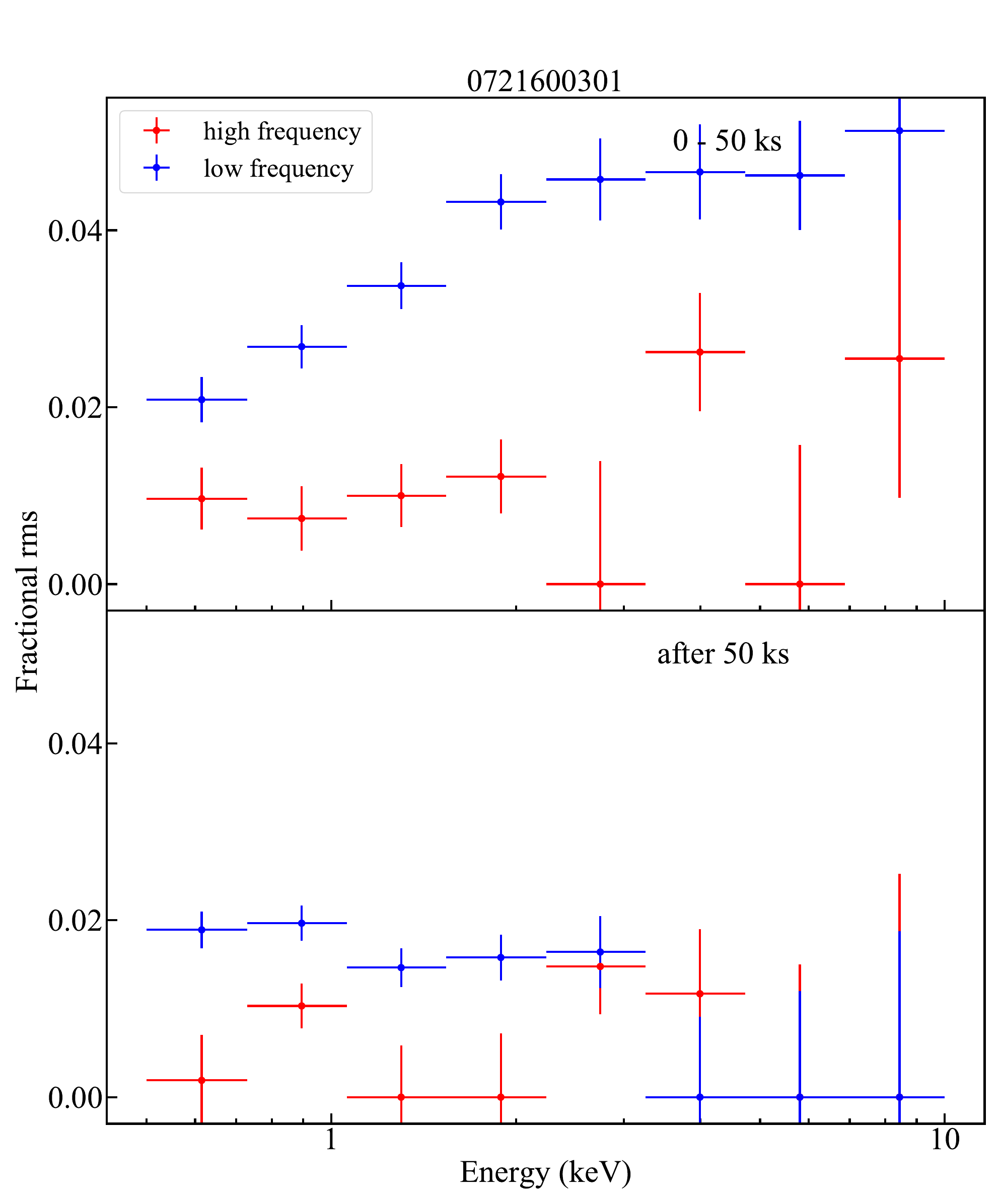}\\
\caption{\label{fig:4thrms} Similar to Fig. \ref{fig:rmsspectra}, but here we present the rms spectra for the 0--50 ks data (upper) and after 50 ks data (lower) for the 4th exposure. The transition from ``harder-when-brighter" to ``softer-when-brighter" seen in Fig.\ref{fig:SR2} is now visible in the low frequency rms spectra (changes from a hard rms spectrum in the upper panel, to a soft one in the lower panel). 
}
\end{figure}  

We have previously demonstrated that the 0.5--10 keV spectral variability of Ark 120 transited from ``harder-when-brighter" to ``softer-when-brighter" during the 4th exposure (Fig. \ref{fig:SR2}). Such a transition however is absent in the spectral variability in 2--10 keV band (Fig. \ref{fig:SR1}). This indicated the abnormal spectral variability could be attributed to the soft X-ray excess component below 2 keV. In Fig. \ref{fig:SR1} we see during the 4th exposure, Ark 120 exhibited no significant spectral variability within 2--10 keV (i.e., with $k$ statistically consistent with zero, and $\chi^2/dof$ $\sim$ 1.0, as labeled in Fig. \ref{fig:SR1}). Assuming no spectral variability of the power-law component, we estimate the expected contribution from the power-law component to the observed 0.5--2.0 keV light curve based on the observed 2--10 keV light curve (using the best-fit spectral model in Fig. \ref{fig:unfolded}). We subtract the power-law contribution, and obtain the light curve for the soft excess component (Fig. \ref{fig:4thLC}). We could clearly see that during the first 50 ks of the 4th exposure, the variation of the soft excess component does not follow that of the power-law. Specifically, while the count rate of the power-law component increases with time, the count rate of the soft excess component remains unchanged or even slightly drops. This could well explain the ``harder-when-brighter" trend within 0.5--10 keV seen during the first 50 ks of the exposure, when the variation is dominated by the (harder) power-law component. After 50 ks, the variation of the soft excess started to follow that of the power-law but with a larger amplitude, the spectral variability thus switched to ``softer-when-brighter". The transition of the spectral variation could also clearly be seen in the different rms spectra between 0--50 ks and after 50 ks (Fig. \ref{fig:4thrms}). Therefore we could conclude that the abnormal X-ray spectral variability during the 4th exposure is due to the variation of the soft excess which could be independent to that of the power-law. Note the light curve of the pure soft excess component (green) in Fig. \ref{fig:4thLC} appears to lag behind the power-law component (yellow), with our CCF analysis yielding a lag of 25.5$^{+3.5}_{-4.0}$ ks. However, this lag is likely unreliable due to the limited duration of the light curves, which spans only about four times the lag. Notably, if we exclude the first 50 ks, the lag disappears entirely. Additionally, this observed lag (25.5 ks) is much larger than the $\sim$900 s soft X-ray lag previously detected in this source \citep{Lobban2018}, further suggesting it may be spurious.

We then examine the abnormal spectral variability in the 6th exposure. This exposure is exceptional in three aspects:

1. It shows a ``harder-when-brighter" pattern within the 2--10 keV range (Fig. \ref{fig:SR1}) but exhibits a very steep ``softer-when-brighter" trend within the 0.5--10 keV range (Fig. \ref{fig:SR2}). 

2. It features a ``V"-shaped rms spectrum with minimum variability around 2 keV (Fig. \ref{fig:rmsspectra}).

3. There is strong coordination between variations in the 2--4 keV and 4--10 keV ranges, but negative coordination between the 0.5--2 keV and 2--10 keV ranges (Fig. \ref{fig:CCF}). 

\begin{figure}
\centering
\includegraphics[width=0.5\textwidth]{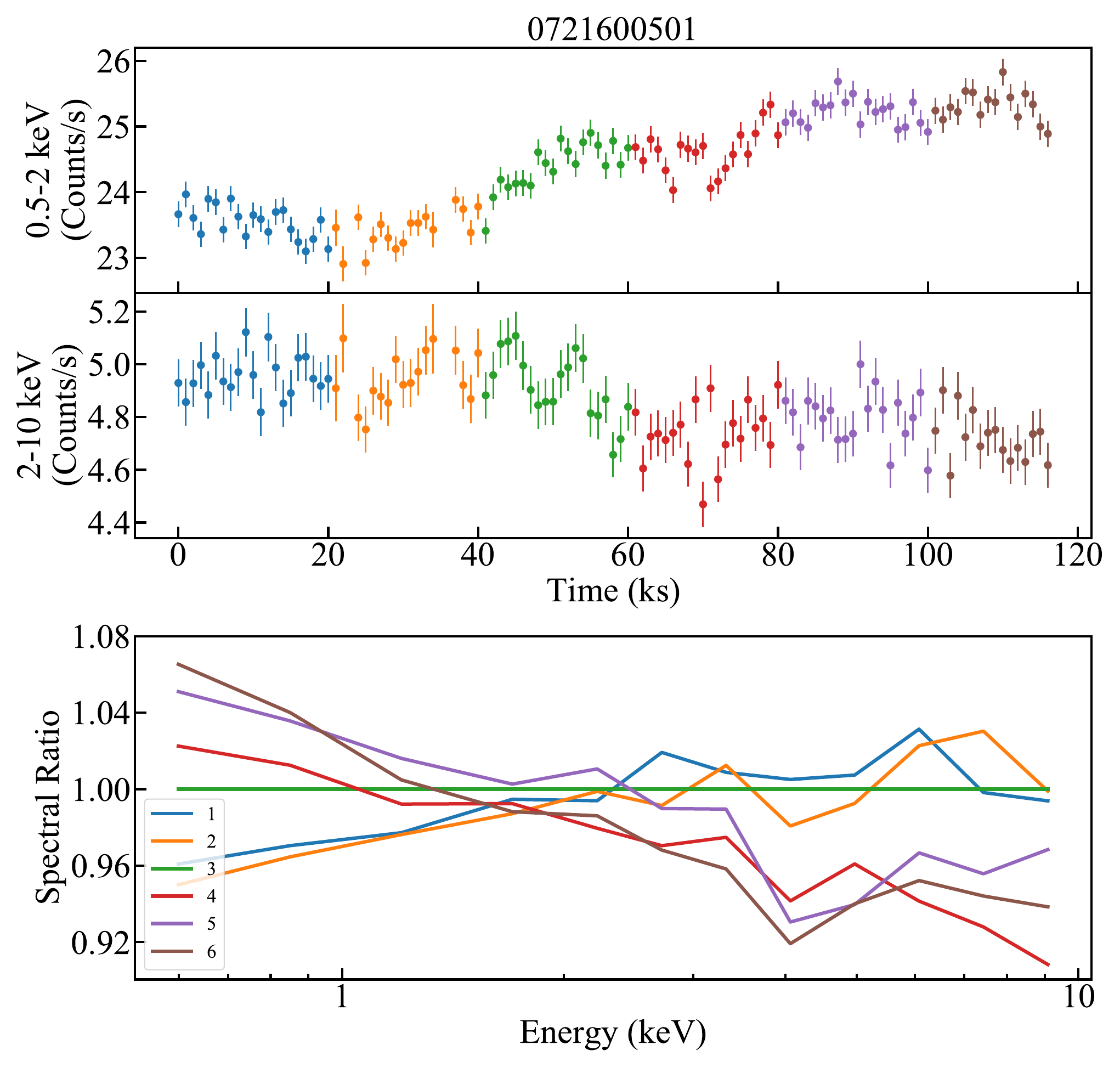}\\
\caption{\label{fig:6th} Upper: we divide the 6th exposure into six equal intervals. Lower: the spectra ratio plots (the ratio of the spectrum from each interval to that of the 3rd interval) pivot at around 2 keV. 
}
\end{figure}

We divide the exposure into six intervals (see Fig. \ref{fig:6th}) and extract their spectra respectively. To visualize the variations between spectra from different intervals, we adopt the spectral ratio method of \cite{Zhang2018}, and plot the ratio of the spectrum from each interval to that of the 3rd interval (Fig. \ref{fig:6th}). In the spectral ratio plot we see power-law-like ratio spectra pivoting at around 2 keV. The slopes of the ratio spectra $<$ 2 keV appear consistent with those at $>$ 2 keV. This indicates the variations of the soft excess well follow those of the power-law component.

The exceptional X-ray spectral variation in this exposure thus could be attributed to the power-law spectrum pivoting at around 2 keV, which could easily explain all the abnormal spectral variability aforementioned: 1) ``harder-when-brighter" within 2--10 keV, but ``softer-when-brighter" within 0.5--10 keV; 2) ``V" shape rms spectrum, and 3) postive coordination between variations in 2--4 and 4--10 keV, but negative coordination between 0.5--2 and 2--10 keV.

We note a pivoting power-law has been adopted to describe the spectral variability of AGNs and X-ray binaries \citep[e.g.,][]{Zdziarski2003}. 

Pivoting at low energy, such as at 2 keV as discovered in this work, however is rare, as Seyfert galaxies generally have a pivot energy $\gg$ 10 keV \citep{Zdziarski2003}. Indeed, ``softer-when-brighter" is naturally expected for a much higher pivot energy. Contrarily, for a much lower pivot energy, we would expect ``harder-when-brighter" within energy range above the pivot energy, and ``softer-when-brighter" below the pivot energy. 
For energy range covering the pivoting energy (such as 0.5--10 keV here), the term ``brighter" becomes misleading as while the flux above the pivot energy increases, the flux below the pivot energy would decrease, thus the spectral variability could not be simply described with ``softer-when-brighter" or ``harder-when-brighter". 

\section{Discussion} \label{sec:discussion}
As introduced in \S\ref{sec:intro}, the nature of the ``softer-when-brighter" trend in Seyfert galaxies is puzzling. At higher X-ray fluxes, more seed photons must have been up-scattered into X-ray. To explain the softer spectrum, the corona has to have either lower temperature, or smaller opacity, or both. Since ``hotter-when-brighter" pattern has been observed in a number of individual sources using NuSTAR spectra \citep{Keek2016,Zhang2018,Kang_2021,Pal2023}, coronal geometry changes must be involved, such as jet-like flares \citep{Wilkins2015}, outflowing corona \citep{Liu2014}, or coronal inflation/contraction \citep{Wu_2020}. But note ``cooler-when-brighter" has also been reported in a few sources \citep{Barua2020,Kang_2021,Wilkins2022}, suggesting a more complicated picture. It is interesting to note that, ``hotter-when-brighter" generally occurs at $\Gamma$ $<$ 2.05, while ``cooler-when-brighter" at $\Gamma$ $>$ 2.05 \citep{Kang_2021}, indicating more than a single underlying mechanism is involved.

Furthermore, \cite{Wu_2020} reported that the ``softer-when-brighter" trend in Seyfert galaxy NGC 4051 is clearly timescale dependent, indicating an empirical ``softer-when-brighter" trend insufficient to explain the observed spectral variability in NGC 4051. A two-tier corona geometry was proposed by \cite{Wu_2020}, in which, coronal inflation, which leads to a smaller opacity of the corona thus softer spectrum, accompanies with long term flux increase, while rapid variations are attributed to small scale flares/nano-flares.

Through analysis of the X-ray spectral variability during six XMM-Newtonexposures of Ark 120 from 2003 to 2014, we find the power-law component within the 2--10 keV range exhibits a clear ``softer-when-brighter" trend on a long timescale of years, but no evident ``softer-when-brighter" trend during individual exposures lasting approximately 120 ks. Therefore, the X-ray spectral variability of the power-law component in Ark 120 is similarly timescale-dependent, further supporting the model of \cite{Wu_2020}. 

Remarkably, a ``harder-when-brighter" trend within 2–10 keV is observed during the 6th XMM-Newtonexposure, indicating that the spectral variability behavior of the power-law component can change over time. Indeed, the spectral variability during the 6th exposure could be described with a power-law pivoting at around 2 keV. Such a low pivot energy is rare for Seyfert galaxies which typically have pivot energy $\gg$ 10 keV \citep{Zdziarski2003}. As the 6th exposure does not have distinct X-ray flux or power-law spectral slope compared with other exposures (Fig. \ref{fig:unfolded}), it is not clear what drives such abnormality. Within the scheme of thermal Comptonization, the output X-ray spectrum is determined by many factors, including the energy and number of seed photons which would be up-scattered, and the geometry, opacity, and temperature of the corona. A pivoting power-law implies a finely tuned situation where when more seed photons are up-scattered into energies above the pivot energy, fewer are up-scattered into energies below the pivot energy. For instance, a pivoting
power-law could be reproduced if the total number of X-ray photons keeps constant, but the opacity of the corona changes (e.g., see Fig. 3 of \citealt{Titarchuk1994}), i.e., higher opacity yields flatter spectrum. Similarly, if the total number of X-ray photons keeps constant, the changes of coronal temperature could also produce a pivoting power-law as hotter corona produces harder spectrum. 

Considering that many factors could have been involved, it is hard to draw a clear physical picture. We note the variation of XMM-OM UVW2 flux of Ark 120 seems to well match the variation of the 0.3--10 keV during this exposure (see Fig. 1 of \citealt{Lobban2018}), suggesting the variation in seed photon flux may have played a role. We also note while two NuSTAR observations on Ark 120 have been obtained, simultaneous to the 2nd and 5th XMM-Newtonexposures, respectively, only lower limits to the high energy cutoff could be derived for the two exposures \citep{Kang2023}, and thus we are unable to examine the variation of coronal temperature in Ark 120.

Nevertheless, the low pivot energy we discovered indicates such an event should be rare, during which the total number of X-ray photons changes little, but the spectral slope changes significantly. This rarity suggests that specific conditions must be met for this phenomenon to occur, possibly involving a delicate balance in the properties of the X-ray corona and seed photon population. Further detailed analysis and comparison with other Seyfert galaxies may help to uncover the underlying mechanisms driving this unusual spectral behavior.\\

Within the 0.5--10 keV range, while a similar ``softer-when-brighter" trend is observed in Ark 120 on long timescales and a weaker trend within individual exposures, there are significant deviations from a single empirical ``softer-when-brighter" trend. Particularly, the spectral variability switches from ``harder-when-brighter" to ``softer-when-brighter" during a single exposure. These observations could be attributed to the contribution of the soft excess component, whose variability sometimes follows and sometimes does not well follow that of the power-law continuum. Note \cite{Nandi2021} reported a strong correlation between the long term variation of the soft excess and the primary power-law continuum, and such a long term correlation is also visible in Fig. \ref{fig:unfolded} where we find strong variations in continuum fluxes between exposures, but weak variations in the relative strength of the soft excess. 
However, during the specific exposure exhibiting transition from ``harder-when-brighter" to ``softer-when-brighter" within 0.5--10 keV, we find the correlation between the soft excess and power-law continuum vanishes in the first 50 ks, and then reappears. 
Assuming the soft excess in Ark 120 originates from a warm corona \citep[e.g.][]{Matt2014,Porquet2018}, our findings highlight the intricate and poorly understood relationship between the variations of the warm corona and the hot corona. A future large-sample study of this correlation is essential to advance our understanding.

\par
Finally, we stress that the X-ray spectral variability in Ark 120 is extremely complex and cannot be described by a single parameter. We have employed various techniques and tools, including multi X-ray band light curves and corresponding softness ratio light curves, softness ratio versus count rate plots, structure functions in various bands and their ratios, rms spectra, cross-correlation functions (CCF), and ratio of X-ray spectra. As demonstrated, none of these techniques alone can fully capture the intricate nature of the variability. Even with all these methods, additional efforts are needed to segment a single exposure into different intervals to understand the variability better.

The variability might be driven by multiple factors, including changes in the accretion rate, variations in the coronal temperature or geometry, and fluctuations in the seed photon population. Such detailed analyses not only highlight the intricate nature of AGN variability but also emphasize the need for multi-faceted approaches in studying these phenomena. Future studies, incorporating even more sophisticated techniques and larger datasets, will be crucial for unraveling the complex mechanisms governing the X-ray variability in Seyfert galaxies like Ark 120.

\section*{Acknowledgement}
We acknowledge the anonymous referee for encouraging comments.
The work is supported by National Natural Science Foundation of China (grant nos. 12033006, 12192221, 123B2042) and the Cyrus Chun Ying Tang Foundations.

The work is based on observations obtained with XMM-Newton, an ESA science mission with instruments and contributions directly funded by ESA Member States and NASA.

\bibliographystyle{aasjournal}
\bibliography{Ark120}{}

\label{lastpage}
\end{document}